\documentclass[10pt, a4paper]{article}

\usepackage{lrec-coling2024} 
\usepackage{microtype}
\usepackage{multirow}
\usepackage{booktabs}
\usepackage{amssymb}
\usepackage{graphicx}
\usepackage{enumitem}
\usepackage{xcolor}

\title{Options-Aware Dense Retrieval for Multiple-Choice query Answering} 

\name{Manish Singh, Manish Shrivastava} 

\address{Language Technologies Research Center, KCIS \\
         IIIT-Hyderabad, Telangana, India \\
         manish.singh@research.iiit.ac.in, m.shrivastava@iiit.ac.in\\}

\abstract{
Long-context multiple-choice question answering tasks require robust reasoning over extensive text sources. Since most of the pre-trained transformer models are restricted to processing only a few hundred words at a time, successful completion of such tasks often relies on the identification of evidence spans, such as sentences, that provide supporting evidence for selecting the correct answer. Prior research in this domain has predominantly utilized pre-trained dense retrieval models, given the absence of supervision to fine-tune the retrieval process. This paper proposes a novel method called Options Aware Dense Retrieval (OADR) to address these challenges. ORDA uses an innovative approach to fine-tuning retrieval by leveraging query-options embeddings, which aim to mimic the embeddings of the oracle query (i.e., the query paired with the correct answer) for enhanced identification of supporting evidence. Through experiments conducted on the QuALITY benchmark dataset, we demonstrate that our proposed model surpasses existing baselines in terms of performance and accuracy. 
 \\ \newline  }

\begin{document}

\maketitleabstract

\section{Introduction}

Multiple-Choice Question Answering (MCQA) is a challenging task in natural language processing , involving understanding and reasoning over the given context to select the right answer from a set of options. While significant progress has been made in MCQA, it becomes more complex when dealing with long-context scenarios, as current language models are limited to processing only a few hundred words at once. In such cases, effective retrieval of relevant evidence spans from the context becomes crucial for accurate answer selection.

Existing approaches in long-context MCQA \cite{bowman2022quality} often employ dense retrieval models \cite{karpukhin2020dense}, which encode a query and context sentences as separate dense representations and employ nearest neighbour search to identify the supporting evidence. The primary objective is to retrieve the most relevant information given a query. However, these methods typically rely on pre-trained retrieval models \cite{karpukhin2020dense} as they lack the supervision required for fine-tuning the retrieval process. Achieving an effective dense representation space often requires a larger number of relevance, which poses a challenge in the context of long-form MCQA where relevance levels for context sentences are not available. Previous studies \cite{bowman2022quality} have consistently demonstrated that when pairing the query with the correct answer (referred to as the oracle query) during the retrieval process, there is a substantial improvement in the quality of the retrieved evidence spans. This improvement in evidence retrieval directly contributes to an overall enhancement in the MCQA model's performance. Consequently, there is a clear demand for innovative techniques to enhance the retrieval stage in long-context MCQA, especially when relevance labels for context sentences are unavailable.

In this research paper, we introduce a novel retrieval method called "Options Aware Dense Retrieval" (OADR) to address the challenges of long-context MCQA. OADR aims to identify crucial evidence spans, like sentences or paragraphs, that support the selection of the correct answer. 

The main concept behind OADR is to enhance the dense retrieval process by fine-tuning it using a generated contrastive dataset.
OADR involves training the model to make query-options embeddings closely resemble those of the oracle query. This encourages the model to prioritize evidence relevant to the correct answer. This approach captures nuanced relationships between the query and answer options, leading to improved evidence identification and better MCQA performance.

 We tested the effectiveness of OADR on the QuALITY benchmark dataset, a respected benchmark for long-context MCQA. Our experiments consistently showed that OADR outperforms existing methods in terms of accuracy and retrieval quality. 
 \begin{figure*}
    \centering
\includegraphics[width=7in, height=2.5in]{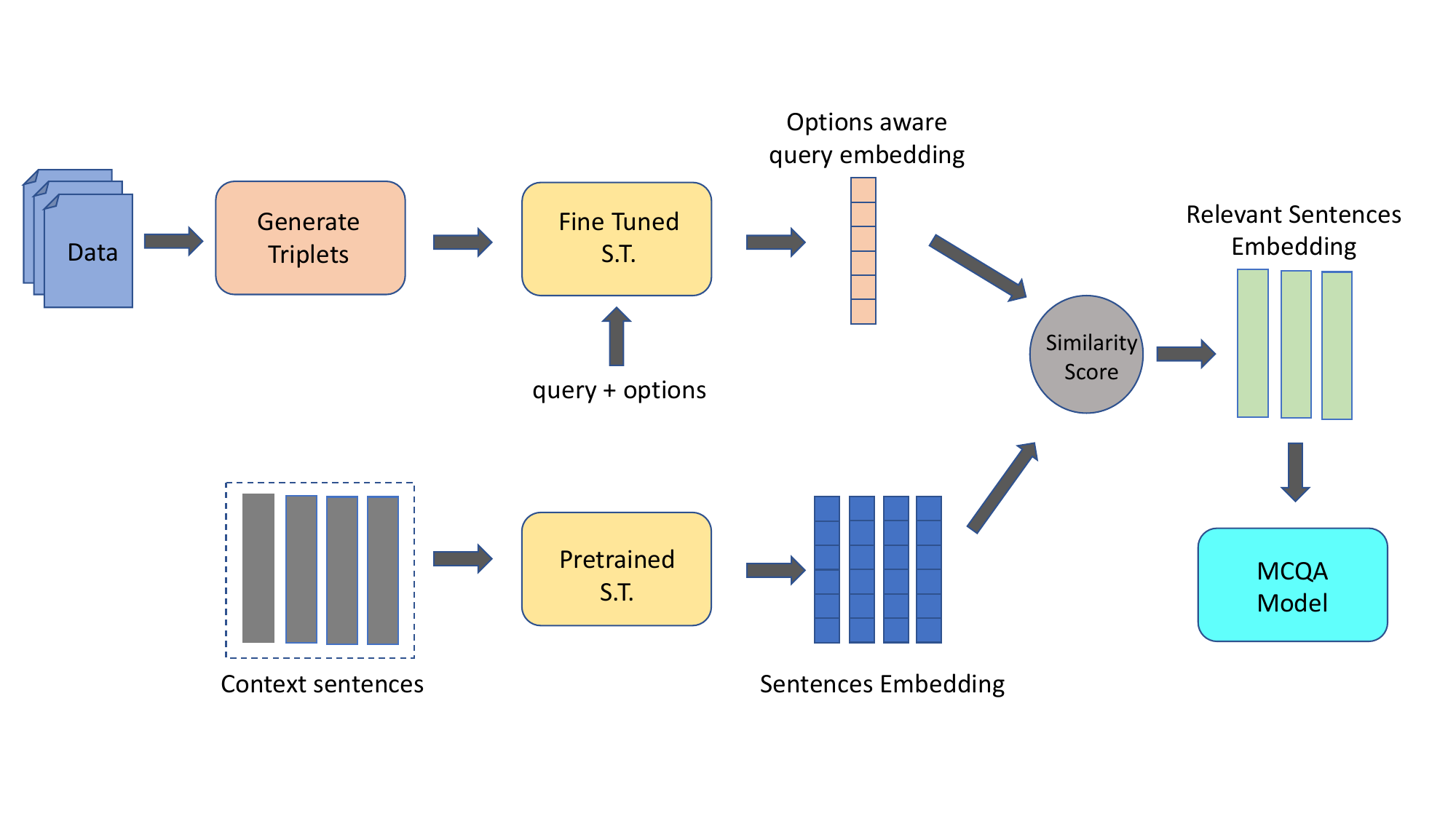}
  \caption{Block Diagram of Options-Aware Dense Retrieval.}
  \label{systemarch}
\end{figure*}

\section{Related Work}

Multiple-choice question answering (MCQA) employs two primary model types: Encoder-Only \cite{devlin2018bert,liu2019roberta,he2021debertav3} and Encoder-Decoder \cite{lewis2019bart,raffel2020exploring}. The former calculates scores based on interactions between context, query, and answer options, selecting the highest-scoring option. In the Encoder-Decoder approach, context, query, and options are combined for answer generation. Existing MCQA datasets \cite{lai2017race,huang2019cosmos,richardson2013mctest} mostly feature shorter contexts, but the QuALITY dataset challenges this norm with an average context length of 5,000 tokens, necessitating dense retrieval for accurate answer selection.

Dense retrieval harnesses models like BERT \cite{devlin2018bert} or RoBERTa \cite{liu2019roberta} to encode queries and contexts, excelling in a dense representation space. While effective in well-labeled scenarios, few-shot performance is less established \cite{xiong2020cmt}. Fine-tuning dense retrieval methods often relies on supervision \cite{zhan2021optimizing, fang2022open}, but long-context MCQA lacks context sentence relevance labels.

Our innovation is inspired by the Setfit \cite{tunstall2022efficient} model, which employs sentence transformers \cite{reimers2019sentence}for few-shot classification. We aim to enhance long-context MCQA retrieval without explicit relevance labels, potentially boosting model performance in relevance level scarce scenarios.

\section{METHODOLOGY} 
To address the long-context MCQA task, we employ a two-step system that includes an extraction step followed by a QA model. The input consists of a query $Q$, a background context $C = <s_1,...,s_M>$ containing M sentences, and a set of $n$ options ${O_1,O_2,...,O_n}$ with only one option being the correct answer. In the extraction step, we identify a set of $K$ evidence spans from the context $C$. These selected sentences are then sorted based on their original order in the passage and concatenated to form a 'passage' specific to that query. In the second step of our approach, we utilize a standard pre-trained language model designed for MCQA to identify the correct answer from the given options. In OADR, our objective is to fine-tune the retrieval process, ensuring that the embeddings of the option-aware query closely resemble those of the oracle query. This fine-tuning enhances the effectiveness of the retrieval stage, increasing the likelihood of correctly identifying the answer. The proposed method is depicted in the block diagram shown in Figure \ref{systemarch}. In the subsequent section, we discuss the steps involved in the method, explaining how each component contributes to the overall approach.

\subsection{Options-Aware Dense Retrieval}

To enable dense retrieval in our approach, we make use of Sentence Transformers. These models employ Siamese and triplet network architectures to generate embeddings for sentences, effectively capturing their semantic meaning. As relevance levels for context sentences are not available, we fine-tuned the Sentence Transformers using a contrastive, Siamese approach on a newly generated dataset.

In this dataset, for each sample, which includes a query (Q), options (O1, ..., On), and the correct answer (O1), we create a triplet (Anchor, Positive, Negative). The Anchor represents the oracle query (Q + O1). The Positive component is formed by concatenating the query with all options (Q + O1 + O2 + O3 + O4). In contrast, the Negative component is created by concatenating the query with all wrong options (Q + O2 + O3 + O4), assuming that O1 is the correct answer.

To fine-tune the Sentence Transformer model, we use this generated dataset and apply TripletLoss during training. TripletLoss works to minimize the distance between the Anchor and Positive sentences while maximizing the distance between the Anchor and Negative sentences. This training objective encourages the Sentence Transformer model to learn embeddings that effectively capture the similarities and differences between relevant and irrelevant options for a given query.

\begin{table*}[tp!]
  \begin{center}
     \begin{tabular}{p{7cm} p{3.5cm} p{4cm}}
    \toprule
      \multirow{2}{*}{\textbf{Model}} & {\textbf{QuALITY}} & \textbf{RACE} $\rightarrow$ \textbf{QuALITY} \\
      & Accuracy & Accuracy \\
    \midrule
    
    Longformer-base & 33.7 / 32.6  & 38.1 / 32.8\\
    LED-base & 25.1 / 24.3 & 38.1 / 32.8 \\
    LED-large & 25.1 / 25.6 & 35.6 / 32.0 \\
      DPR -> RoBERTa-base & 40.0 / 36.4 & 43.8 / 37.2\\ 
      DPR -> RoBERTa-large & 26.7 / 24.0 & 50.8 / 46.2\\
      DPR -> DeBERTaV3-base & 41.8 / 37.4 & 46.7 / 40.9\\
      DPR -> DeBERTaV3-large & 45.1 / 39.2 & 53.6 / 47.4 \\
      ST(pre-train) -> RoBERTa-base & 39.4 / 35.3 & 43.1 / 37.4\\
      ST(pre-train) -> DeBERTaV3-base & 40.6 / 38.1 & 44.6 / 39.5 \\
      OADR -> RoBERTa-base & 43.5 / 36.8  & 47.2 / 41.4\\ 
      OADR -> RoBERTa-large & 32.4 / 27.7 & 53.7 / 46.8\\
      OADR -> DeBERTaV3-base & 44.1 / 38.5 & 51.6 / 43.2\\
      OADR -> DeBERTaV3-large & 49.2 / 42.4 & 59.3 / 48.9\\

    \bottomrule

    \end{tabular}
    \caption{Accuracy on full QuALITY and QuALITY-HARD subset(full/HARD).}
    \label{tab:table1}
  \end{center}
\end{table*}

\subsection{MCQA Model} 
In our approach, instead of feeding the entire context into the MCQA model or truncating it, we leverage the fine-tuned OADR to extract the most relevant sentences from the context. This significantly improves the model's ability to identify the correct answer, particularly in long-context scenarios where traditional methods may face limitations. As a result, we can employ high-performing short-sequence encoder models like RoBERTa, DeBERTaV3, or encoder-decoder models like T5\cite{raffel2020exploring} and BART \cite{lewis2019bart} as our MCQA model.

\textbf{Inference} 
During inference, we use a specific procedure. We begin by creating an options-aware query that combines the query with each option and pass it through a fine-tuned sentence transformer to obtain an options-aware query embedding. Context sentences are processed individually using a pre-trained sentence transformer to generate dense representations. Relevant sentences are chosen by calculating negative Euclidean (L2) distances from the options-aware query embedding. After identifying the relevant sentences, we sort them based on their original order and concatenate them to form a coherent passage. The length of the passage is limited to a maximum of 300 tokens, ensuring concise and manageable input for downstream tasks. Finally, this constructed passage is fed into the trained MCQA model, which considers the options and the context to determine the most suitable answer among the given options.

\section{Experiments}
In this section, we cover three main aspects: the corpora used for training and evaluation, the comparison baselines, and our approach's implementation details. 

\subsection{Dataset} 

The QuALITY benchmark, focusing on long-context MCQA, comprises two subsets: QuALITY-EASY and QuALITY-HARD, representing varying query difficulty levels. The context passages in the benchmark are about 5,159 tokens long, with 2,523 training queries. Both validation and test set contain around 2000 samples.  To improve our training process, we included additional MCQA data from the RACE\cite{lai2017race} dataset, which has shorter passages but a substantial number of queries, totaling around 88,000. This presents an advantageous opportunity for knowledge transfer, allowing us to enrich our model. We used the entire RACE dataset to train the OADR and the MCQA model.

\subsection{Baseline}
Our comparison includes two sets of baselines: those from published research and our own implementations. The former includes the Longformer \cite{beltagy2020longformer}encoder model, accommodating up to 4,096 tokens, the Longformer Encoder-Decoder (LED) model with an extended 16,000-token limit, and the DPR \cite{karpukhin2020dense} model for open-domain retrieval. We also introduced a pre-trained Sentence Transformer for sentence extraction, coupled with an encoder-based MCQA model for answer selection. Notably, the queries given to these models remain in their original, unaltered form.

\subsection{Implementation Details}
We utilized the "multi-qa-mpnet-base-dot-v1" model as the sentence transformer for our retrieval process. The retrieval model was fine-tuned with a learning rate of 1e-4, using a batch size of 8, and a maximum sequence length of 128 tokens for one epoch. For answer selection, we employed RoBERTa and DeBERTaV3 as our MCQA models, both fine-tuned with a learning rate of 5e-4. We used a batch size of 4 for the base-model and a batch size of 2 for the large-model.

\begin{table}[h!]
{
  \begin{center}
    \begin{tabular}{p{5cm} p{1.5cm}}
    \toprule
     Models  & EM  \\
     \midrule
     $Q$ & 53.6   \\
     $OAQ$ & 50.3  \\
     $(OAQ)_R$& 61.4 \\ 
     $OAQ_{RQ}$ & 62.1 \\
    
    \bottomrule
    \end{tabular}
    \caption{Average \% over-lap of retrieved sentences with sentences retrieved sentences from $OQ$ on dev-set}
    \label{tab:overlap}
  \end{center}
}  
\end{table}

\section{Results and Analysis}
This section compares our system's performance against different baselines, employing accuracy as the primary evaluation metric. Additionally, we delve into an analysis of the learned options-aware query embeddings, providing empirical evidence in tables and plots.
\subsection{Overall Result} 

Table \ref{tab:table1}  provides a performance comparison between our OADR model and various baseline models. The results clearly highlight that the OADR model with the DeBERTaV3-large variant achieves the highest performance among all examined baseline models. Notably, when comparing models trained with different datasets, it is evident that the combination of the RACE dataset followed by the QuALITY dataset (RACE -> QuALITY) leads to significantly better performance than models trained exclusively on the QuALITY dataset. This performance improvement is attributed to the relatively limited size of the QuALITY training set, underscoring the benefit of knowledge transfer from the larger and more diverse RACE dataset.

\subsection{Learned Options-Aware Query Representation}

\begin{figure}
    \centering
    \includegraphics[width=7cm, height=5cm]{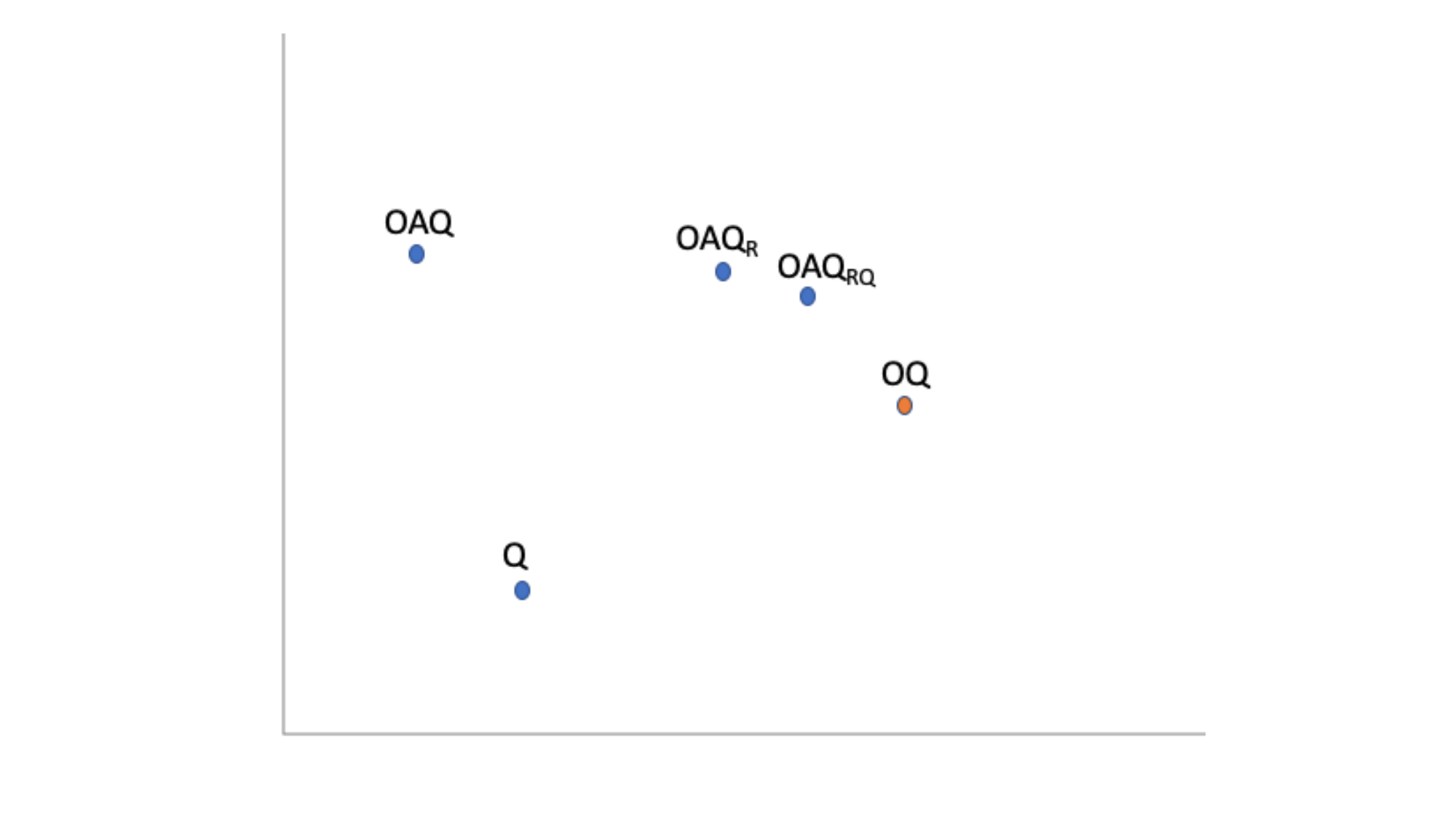}
    \caption{t-SNE plot for different queries}
    \label{fig:t-sne}
\end{figure}

In an assessment to determine if OADR can replicate oracle query embeddings, we selected a random example from the development set (question-unique-id: 51650-RM2TQ88X-1). Using t-SNE \cite{van2008visualizing} for visualization, various query embeddings were generated as shown in Figure \ref{fig:t-sne}, including the query embedding $Q$ from a pre-trained Sentence Transformer, the oracle query embedding $OQ$ from the same pre-trained model, the options-aware query embedding $OAQ$  from the pre-trained Sentence Transformer, and two variations of options-aware query embeddings from OADR, one fine-tuned on RACE $OAQ_R$ and the other fine-tuned on both RACE and QuALITY $OAQ_{RQ}$.

The visualizations showed that the options-aware query embedding $OAQ$ was the farthest from the oracle query embedding $OQ$, while the fine-tuned options-aware query embeddings ($OAQ_R$ and $OAQ_{RQ}$) were positioned closest to the oracle query embedding. This observation strongly suggests that options aware query from OADR effectively mimics the embeddings of the oracle query.

In the second experiment, we used sentences retrieved from the oracle query as the benchmark for retrieval performance and calculated the percentage overlap with sentences obtained from different query embeddings. The results in Table 2 revealed that concatenating all the options with the query for retrieval led to a decrease in performance. However, a substantial increase in overlap was observed when utilizing the options-aware query embedding derived from OADR fine-tuned on RACE. This finding strongly indicates that OADR places a higher priority on evidence sentences highly relevant to the correct answer, resulting in improved retrieval performance.

\section{Conclusion}
The study introduces Options Aware Dense Retrieval (OADR) to enhance MCQA in long-context scenarios by fine-tuning dense retrieval with a contrastive dataset. OADR uses query-options embeddings to prioritize relevant evidence. It outperforms existing methods in accuracy and retrieval quality on the QuALITY benchmark. Future research can explore further refinements, including different contrastive dataset constructions and additional contextual information integration, for enhanced retrieval accuracy.


\label{lr:ref}
\bibliographystyle{lrec-coling2024-natbib}
\bibliography{lrec-coling2024-example}


\end{document}